# Ultrafast terahertz detectors based on three-dimensional meta-atoms

B. Paulillo[1,†], S. Pirotta[1,†], H. Nong[2], P. Crozat[1], S. Guilet[1], G. Xu[1,3], S. Dhillon[2], L. H. Li[4], A.G. Davies[4], E.H. Linfield[4], And R. Colombelli[1,*]

[1]*Centre de Nanosciences et de Nanotechnologies, CNRS, Univ. Paris-Sud, Université Paris-Saclay, C2N – Orsay, 91405 Orsay cedex, France*
[2]*Laboratoire Pierre Aigrain, Département de physique de l'ENS, École normale supérieure, PSL Research University, Université Paris Diderot, Sorbonne Paris Cité, Sorbonne Universités, UPMC Univ. Paris 06, CNRS, 75005 Paris, France*
[3]*Key Laboratory of Infrared Imaging Materials and Detectors, Shanghai Institute of Technical Physics, Chinese Academy of Sciences, Shanghai 200083, China*
[4]*School of Electronic and Electrical Engineering, University of Leeds, Woodhouse Lane, Leeds LS2 9JT, United Kingdom*
*Corresponding author: raffaele.colombelli@u-psud.fr*

**ABSTRACT**

Terahertz (THz) and sub-THz frequency emitter and detector technologies are receiving increasing attention, underpinned by emerging applications in ultra-fast THz physics, frequency-combs technology and pulsed laser development in this relatively unexplored region of the electromagnetic spectrum. In particular, semiconductor-based ultrafast THz receivers are required for compact, ultrafast spectroscopy and communication systems, and to date, quantum well infrared photodetectors (QWIPs) have proved to be an excellent technology to address this given their intrinsic ps-range response However, with research focused on diffraction-limited QWIP structures ($\lambda/2$), RC constants cannot be reduced indefinitely, and detection speeds are bound to eventually meet un upper limit. The key to an ultra-fast response with no intrinsic upper limit even at tens of GHz is an aggressive reduction in device size, below the diffraction limit. Here we demonstrate sub-wavelength ($\lambda/10$) THz QWIP detectors based on a 3D split-ring geometry, yielding ultra-fast operation at a wavelength of around 100 µm. Each sensing meta-atom pixel features a suspended loop antenna that feeds THz radiation in the ~20 µm³ active volume ($V_{eff} \sim 3\times10^{-4}$ $(\lambda/2)^3$). Arrays of detectors as well as single-pixel detectors have been implemented with this new architecture, with the latter exhibiting ultra-low dark currents below the nA level. This extremely small resonator architecture leads to measured optical response speeds - on arrays of 300 devices - of up to ~3 GHz and an expected device operation of up to tens of GHz, based on the measured S-parameters on single devices and arrays.

## 1. INTRODUCTION

Applications of terahertz (THz) frequency radiation (loosely defined to span the 1–10 THz frequency region) range from medical imaging through to trace gas spectroscopy. Each particular application poses specific requirements on the system components, but ultra-fast operation—for either emission or detection—is usually a vital functionality (e.g. in wireless communications, or in real-time imaging). To date, many THz detectors (bolometers, pyroelectric crystals, Golay cells) contain thermal sensing elements that yield modest detectivities ($D^* \sim 10^9$ cm√(Hz/W)) and/or require cryogenic cooling. Furthermore, their bandwidth is normally limited to tens of kHz by the device thermal and/or electrical time constants ($\tau \sim$ ms) [1]. Recently, superconducting hot-electron bolometer (SHEB) technology has enabled sensitive and fast responses, which suits niche applications (e.g. heterodyne receivers for astronomy). The best available NbN SHEBs exhibit high responsivities ($\approx$kV/W) and fast response times ($\approx$50 ps) up to 6 THz [2]. For lower frequencies, electronic circuit detectors (e.g. Schottky diodes or field-effect transistors) are in general excellent solutions in the sub-THz frequency range [3]. Renewed interest in compact, semiconductor-based ultra-fast THz detectors has recently appeared, underpinned in part by applications in ultra-fast THz physics [4, 5], frequency-combs technology [6, 7] and pulsed laser development [8–10]. For instance, graphene and related 2D materials are extensively investigated as sensing elements for THz radiation, owing to their ultra-fast carrier dynamics and possible integration with the silicon platform [11]. Recently fabricated devices show a 50 ps rise time, although a responsivity only of the order of nA/W was reported [12].

To date, the most reliable THz receivers combining ultra-fast response (bandwidth in the tens of GHz) and good responsivity ($\approx$A/W) exploit a mature, solid-state technology such as the quantum-well infrared detector (QWIP) [13]. In the mid-infrared (8 µm<$\lambda$<12 µm), QWIPs are at an advanced development stage and used as single-element detectors and as imaging cameras. In the THz range (2-5 THz), QWIP technology is limited to low temperatures, with typical background limited infrared temperatures ($T_{blip}$) of ≈15/20 K [14]. The material system employed is usually GaAs/AlGaAs, although interest has arisen recently in the GaN platform [15]. QWIP technology is promising for efficient THz photon detectors with an ultra-fast response, owing to the short lifetime of intersubband (ISB) transitions ($\tau_{ISB}$ ≈ ps). Indeed when the material carrier dynamics is intrinsically ultra-fast, the only limitation to the detector frequency response arises from the device size i.e. the electrical RC constant of the optoelectronic component plus the read-out circuit. In a very recent paper H. Li et co-workers report a measured bandwidth of 6.2 GHz for a THz QWIP processed in a 400x400 µm$^2$ mesa geometry [16]. This demonstration has been possible exploiting the beating between Fabry-Perot modes of a 6-mm-long THz quantum cascade laser (QCLs). Before that, the best measured speeds were limited to the MHz/sub-MHz range [17], [18]. However, to further reduce the RC time constant and develop an architecture with the potential of response bandwidths in the tens of GHz at THz frequencies, an ultra-small, micron-sized resonator geometry is crucial. Unlike in the NIR/MIR range [19], this is challenging because it means operating in the extreme sub-wavelength limit. (Constraints exist linked to the lack of readily-available THz QC lasers with fast modulation capabilities, but this is beyond the scope of this paper).

To date, THz QWIPs have mainly been embedded in large (compared with the wavelength) photonic structures such as mesas (normally substrate-coupled through a polished facet)[13, 15] or metal diffraction gratings [20]. More recently, arrays of diffraction-limited (λ/2) patch antennas [21] and planar meta-material surfaces [22] have been used to couple radiation into THz quantum well detectors. Indeed, meta-materials are appealing as in-coupler elements for ultrafast and small THz sensors. They rely on extremely sub-wavelength single units called meta-atoms, whose archetypical example is the split-ring resonator: a sub-wavelength LC circuit where the microscopic capacitor can, in principle, host a semiconductor active region. Passive devices (absorbers, polaritonic structures, 3D metamaterials) based on this concept have been demonstrated [23–25]. However, an electrically-injected optoelectronic 'meta-atom' device (a laser or a detector) has yet to be demonstrated. Beside the intrinsically small RC constant, the resonant LC geometry is appealing since it provides new degrees of freedom with respect to conventional photonic cavities: loss and radiation in-coupling can be engineered *via* critical coupling/free space impedance matching; frequency tuning *via* external lumped elements; choice of the absorption directivity *via* antenna design, etc.

## 2. DEVICE CONCEPT, MODELING AND FABRICATION

Here, we report meta-atom QWIP detectors with dimensions below the diffraction limit (λ/2) operating around 3 THz with an ultra-fast response. The device concept is illustrated in Figure1(a).

The device core is a sub-wavelength-sized metal-semiconductor-metal (MSM) capacitor, shown in blue in Figures 1 (a) and (b), which hosts the detector active region, connected to a suspended metallic loop antenna providing inductance. We have shown in [24] that a similar object behaves as an LC circuit, where the capacitance and the inductance are set by the geometry. Owing to this circuital character, the MSM capacitor can be extremely sub-wavelength, while still permitting an electromagnetic (EM) resonance. However, electrical injection in this system has proved to be challenging. Here, we have finally solved this problem by patterning the ground plane into two separated regions, with the gap between the two defining an additional capacitance ($C_{gap}$). This forces the DC detector current to flow through the active core, while $C_{gap}$ only marginally affects the electromagnetic resonance in the THz range. The equivalent circuit is shown in Figure 1(b). Figure 1(c) shows optical microscope and SEM images of an array of devices, as well as an SEM image of a single detector pixel.

To highlight the key role of the small loop antenna as a receiving element for THz radiation, the optical response of the structure under illumination with a normally incident plane wave has been simulated with a commercial finite element solver (Comsol Multiphysics). The results are shown in Figure 2.

When the antenna is not present and the sub-wavelength MSM capacitor alone is considered, a Fano-like feature appears in the reflectivity spectrum around 3.5 THz (Figure 2, dashed-dot grey line, top panel): it arises from the Wood's anomaly of the grating created by the sub-wavelength openings in the ground plane [26, 27]. Importantly, the ohmic dissipation in the MSM cavity as a function of excitation frequency reveals that no energy is dissipated in the active region: the MSM capacitor size is much smaller than the wavelength, hence no optical mode is available for coupling. It is instead the suspended antenna that enables THz radiation coupling in the full device. When the antenna is present, the electronic LC resonance appears and yields the resonant response shown by the black solid line in Figure 2 (top panel). Energy is now resonantly dissipated into the active region (Figure 2, bottom panel). Note: the hybrid mode resulting from the coupling of the LC mode with the grating results in a larger resonant absorption bandwidth (3 to 4.5 THz), which is beneficial for detector operation.

Driven by these results, arrays and individual meta-atom detectors have been fabricated using standard micro-fabrication techniques (Figure 1(c)) [see Supplementary Material for further details]. The diameter of the active region is only 4 µm, i.e.

only λ/25 for operation between 2 and 4 THz. The surface area of the array is about 350 x 300 μm$^2$ and contains 300 single elements. The semiconductor material used in this work (wafer L1258) is a copy of the structure V267 proposed and extensively studied in [14]. It presents an operating window of about 1.5 THz between 2 and 3.5 THz as confirmed by a preliminary characterization in a standard mesa geometry [see Supplementary Material].

## 3. OPTICAL AND ELECTRICAL CHARACTERIZATION

A complete optical and electrical characterization of the fabricated samples has been performed [details of the experimental arrangement are given in Supplementary Material].

Figure 3(a) shows the low temperature (4.5 K) photocurrent spectrum of an array of meta-atom detectors at different polarizations of the incident beam. As expected from the literature, as well as from the mesa characterization, the spectrum is centered around 3 THz. A clear polarization-sensitive response is observed: maximum coupling is achieved when the magnetic field of the incident beam is orthogonal to the antenna (red curve), confirming the numerical results presented in Figure 2. A residual coupling in the other polarization (blue curve) is probably due to the non-zero $E_z$ component, given by the finite numerical aperture of the lens focusing the incident beam. Furthermore, we have measured the photocurrent spectra from several single meta-atom detectors, individually contacted. They show a response similar to the array configuration (inset of Figure 3(a)), thus confirming that the device arrays operate as optoelectronic meta-surfaces, where the total response is the (incoherent) sum of the individual meta-atoms contributions.

Figure 3(b) shows the device dark current at 4.5 K (black curve), and under illumination from a 300 K blackbody (red curve, more details in Supplementary Material). The operating current densities are in agreement with this type of QWIP design. However, owing to the extremely small surface of the detector active region, the operating currents, have record low values in the 1-10 nA range. For instance, in these devices, the combined semiconductor surface in a 350x300 μm$^2$ array is only 3.6%. The background-limited infrared photo-detection temperature ($T_{BLIP}$) of the devices is 8 K, similar to the temperature found when the same material is processed in a standard mesa detector configuration. The great reduction of the operating dark current enabled by our structures is not accompanied by an appreciable improvement of $T_{BLIP}$. A basic parameter to quantify the response of a photo-detector is the responsivity, defined as $R=(I_{BB@300K}-I_{dark})/P_{incident}$, where $P_{incident}$ is the power impinging on the device array estimated in about 20 nW (see Supplementary Material). It can be readily evaluated using the Planck's law for a 300 K ideal blackbody upon integration of the spectral radiance over the detector sensitivity window previously measured. From the dark/light characterizations we can extract the responsivity of an array of meta-atom QWIPs, as shown in Figure 3(c). The experimental values are in agreement with the literature, both for a detector processed as a large mesa [14] or as small patch cavities [21]. We can conclude therefore that the massive reduction in size of the detector preserves the performance of the QWIP material.

## 4. RF RESPONSE

We now prove that these detectors exhibit ultra-fast speed. First, we performed a direct optical response analysis, using an RF-modulated THz quantum cascade laser (QCL) as the source, and connecting the amplified output of a meta-atom detector array to a spectrum analyzer (SA), as shown in Figure 4(a).

The laser used is a single-mode 3 THz surface-emitting QCL [28], biased in continuous-wave (CW) just under threshold. The signal from a RF synthesizer (13 dBm) is injected using a bias-T. The effect is to switch the laser on and off at the RF frequency. Figure 4(b) shows the QWIP signal, detected after amplification, when the THz laser is modulated at 1.5 GHz. The noise floor is at –82 dBm (laser shielded) for a resolution bandwidth of 1 kHz. With the laser on the detector, the SA reveals a signal at the RF modulation frequency with ~4 dB intensity above the noise. Finally, Figure 4(c) shows the SA signal for RF modulation frequencies of 0.5 (13 dBm), 1.5 (13 dBm) and 2.5 GHz (15 dBm). This demonstrates that the meta-atom detector can operate at speeds at least up to 2.5 GHz.

To gain a deeper insight into the high-frequency response of this detector architecture, we have measured the complex one port S-parameters (real and imaginary part of $S_{11}$) of several single-pixel and arrays of detectors, using a 40-GHz network analyzer.

In Figure 5(a) the data for a representative single device (red curve, top panel) and an array (red curve, bottom panel) are plotted as normalized impedances $z=Z_L/Z_0$ ($Z_L$ is the device impedance loading the $Z_0$=50 Ω measurement line) on a Smith chart. Note: the measurements were made at room temperature; the Smith chart summarizes the real and imaginary part of the $S_{11}$ parameter on a single plot.

The Smith plot confirms that, as expected, the device sub-wavelength size prevents a good impedance-matching with the 50 Ω line over the entire frequency range.

We have inferred the equivalent RF circuit of each device by fitting the predicted behavior (blue curves) against the experimental data (red curves) using commercial circuit simulation software (Keysight ADS). Figure 5(b) shows the equivalent circuit of a single meta-atom detector, including the access contact lines. The circuit modeling the actual detector (within the dashed brown line) comprises the parallel combination of the capacitance $C_c$ and resistance $R_c$ that models the Au-Ti/n-GaAs Schottky contact situated on top of the MSM capacitor. This junction is in series with a resistor $R_{AR}$ accounting for the resistance of the semiconductor active core, and the resistance of the bottom ohmic Pd-Ge contact. The geometric capacitance of the MSM cavity $C_{AR}$ appears in parallel across the contacts while the inductive behavior of the current-carrying metallic parts (including the small loop THz antenna) is modeled with a series inductor L. Finally, $R_a$ is the access resistance and $C_a$ is the capacitance of the large bonding pad. The averaged values extracted from five devices are reported in the inset to Figure 5(b). $C_{AR}$ and $C_c$ are comparable with the estimate based on analytical formulas for the geometric and Schottky capacitance ($C_{AR}$~ 1 fF and $C_c$ ~ 50 fF). Also, the extracted series resistance $R_{AR}$ is of the same order of magnitude as the one extracted from the DC I-V curves, (typically a few kΩ). This model fits the data well (blue and red curves in Figure 5(a)) both for a single pixel and an array of devices. Note: in the latter case the equivalent circuit is just a 2D array of single-pixel equivalent circuits, with small parasitic L and C added between neighboring elements to account for mutual coupling.

The excellent agreement between predicted and experimental values suggests the use of the equivalent circuit as a predicting tool for the detector high-frequency behavior. In particular, the RF signal extraction efficiency can be simulated over a broad frequency range: if a (photo)current generator is added in series to $R_{AR}$, the ratio between the extracted current ($I_{out}$) and generated photocurrent in the active region ($I_{in}$) as a function of frequency can be calculated. The result is shown in Figure 5(c): the continuous black curve represents the frequency response of a single pixel from 100 MHz to 1 THz using the R,L,C values of panel (b), that correspond to the present experimental configuration. The modeling reveals that the Schottky capacitance $C_c$ limits the low frequency current extraction, while the high-frequency cut-off - that in these devices is tens of GHz - is only limited by the pad capacitance $C_a$. As expected, extremely sub-wavelength detectors lead to high-frequency operation, possibly reaching the upper bandwidth limit imposed by the QWIP transport mechanisms (BW≈ 30 GHz if we assume $\tau_{photocarrier} \approx$ 5 ps [29, 30].

## 5. CONCLUSIONS

In conclusion, we have demonstrated terahertz detectors based on a 3D meta-atom geometry with ultrafast response. We have exploited an LC approach to drastically reduce the effective detector area, going beyond the intrinsic limitation given by the RC constant of the diffraction-limited THz structures such as mesas or patch antennas.

Future developments will focus on achieving better response at low response frequencies, replacing the top Schottky contact with an alloyed ohmic contact (Figure 5(c) continuous grey curve). Additionally, a further reduction of the contact pad capacitance (e.g. by a factor 10, Figure 5(c) dashed grey curve) can push the RC high-frequency cutoff to higher frequency. This is of interest in view of using these detectors in combination with mode-locked THz (QC) lasers [8,9,31], that is another important vista of this work. Finally, the extraction efficiency of the RF signal is only a few percent due to the impedance mismatch arising from the very small detector size and the read-out line. This is a known problem in RF technology. It can be overcome, over a small bandwidth around a chosen precise modulation frequency [32], by impedance-matching to 50 Ohm using discrete reactive components (inductors, capacitors).

**Funding**. We acknowledge financial support from the ERC "GEM" grant (Grant Agreement No. 306661), the EPSRC grants 'COTS' and 'HyperTerahertz' (EP/J017671/1 and EP/P021859/1 ), the FET-Open ULTRAQCL 665158 grant. This work was partly supported by the French RENATECH network. We also acknowledge support from the Royal Society and Wolfson Foundation. G.X. acknowledges support from the National Natural Science Foundation of China under Grant No. 61574149, and "The Hundred Talents Program" of CAS.

**Acknowledgment**. We thank A. Degiron, F. Julien, S. Barbieri and A. Delga for useful discussions.

[†]These authors contributed equally to the work.
See Supplementary material

# REFERENCES

1. F. Sizov and A. Rogalski, "THz detectors," Prog. Quantum Electron. **34**, 278–347 (2010).
2. I. Milostnaya, A. Korneev, M. Tarkhov, A. Divochiy, O. Minaeva, V. Seleznev, N. Kaurova, B. Voronov, O. Okunev, G. Chulkova, K. Smirnov, and G. Gol'tsman, "Superconducting Single Photon Nanowire Detectors Development for IR and THz Applications," J. Low Temp. Phys. **151**, 591–596 (2008).
3. A. Lisauskas, S. Boppel, M. Mundt, V. Krozer, and H. G. Roskos, "Subharmonic Mixing With Field-Effect Transistors: Theory and Experiment at 639 GHz High Above fT," IEEE Sens. J. **13**, 124–132 (2013).
4. D. Dietze, J. Darmo, and K. Unterrainer, "Efficient population transfer in modulation doped single quantum wells by intense few-cycle terahertz pulses," New J. Phys. **15**, 65014 (2013).
5. B. Green, S. Kovalev, V. Asgekar, G. Geloni, U. Lehnert, T. Golz, M. Kuntzsch, C. Bauer, J. Hauser, J. Voigtlaender, B. Wustmann, I. Koesterke, M. Schwarz, M. Freitag, A. Arnold, J. Teichert, M. Justus, W. Seidel, C. Ilgner, N. Awari, D. Nicoletti, S. Kaiser, Y. Laplace, S. Rajasekaran, L. Zhang, S. Winnerl, H. Schneider, G. Schay, I. Lorincz, A. A. Rauscher, I. Radu, S. Mahrlein, T. H. Kim, J. S. Lee, T. Kampfrath, S. Wall, J. Heberle, A. Malnasi-Csizmadia, A. Steiger, A. S. Muller, M. Helm, U. Schramm, T. Cowan, P. Michel, A. Cavalleri, A. S. Fisher, N. Stojanovic, and M. Gensch, "High-Field High-Repetition-Rate Sources for the Coherent THz Control of Matter," Sci Rep **6**, 22256 (2016).
6. M. Rösch, G. Scalari, M. Beck, and J. Faist, "Octave-spanning semiconductor laser," Nat. Photonics **9**, 42–47 (2014).
7. D. Burghoff, T.-Y. Kao, N. Han, C. W. I. Chan, X. Cai, Y. Yang, D. J. Hayton, J.-R. Gao, J. L. Reno, and Q. Hu, "Terahertz laser frequency combs," Nat. Photonics **8**, 462–467 (2014).
8. S. Barbieri, M. Ravaro, P. Gellie, G. Santarelli, C. Manquest, C. Sirtori, S. P. Khanna, E. H. Linfield, and A. G. Davies, "Coherent sampling of active mode-locked terahertz quantum cascade lasers and frequency synthesis," Nat. Photonics **5**, 306–313 (2011).
9. F. Wang, K. Maussang, S. Moumdji, R. Colombelli, J. R. Freeman, I. Kundu, L. Li, E. H. Linfield, A. G. Davies, J. Mangeney, J. Tignon, and S. S. Dhillon, "Generating ultrafast pulses of light from quantum cascade lasers," Optica **2**, 944 (2015).
10. F. Wang, H. Nong, T. Fobbe, V. Pistore, S. Houver, S. Markmann, N. Jukam, M. Amanti, C. Sirtori, S. Moumdji, R. Colombelli, L. Li, E. Linfield, G. Davies, J. Mangeney, J. Tignon, and S. Dhillon, "Short Terahertz Pulse Generation from a Dispersion Compensated Modelocked Semiconductor Laser," Laser Photon. Rev. **11**, 1700013 (2017).
11. F. H. L. Koppens, T. Mueller, P. Avouris, A. C. Ferrari, M. S. Vitiello, and M. Polini, "Photodetectors based on graphene, other two-dimensional materials and hybrid systems," Nat. Nanotechnol. **9**, 780–793 (2014).
12. M. Mittendorff, S. Winnerl, J. Kamann, J. Eroms, D. Weiss, H. Schneider, and M. Helm, "Ultrafast graphene-based broadband THz detector," Appl. Phys. Lett. **103**, 21113 (2013).
13. H. Schneider and H. C. Liu, *Quantum Well Infrared Photodetectors: Physics and Applications*, Springer Series in Optical Sciences (Springer, 2007).
14. H. Luo, H. C. Liu, C. Y. Song, and Z. R. Wasilewski, "Background-limited terahertz quantum-well photodetector," Appl. Phys. Lett. **86**, 1–3 (2005).
15. H. Durmaz, D. Nothern, G. Brummer, T. D. Moustakas, and R. Paiella, "Terahertz intersubband photodetectors based on semi-polar GaN/AlGaN heterostructures," Appl. Phys. Lett. **108**, 201102 (2016).
16. H. Li, W.-J. Wan, Z.-Y. Tan, Z.-L. Fu, H.-X. Wang, T. Zhou, Z.-P. Li, C. Wang, X.-G. Guo, and J.-C. Cao, "6.2-GHz modulated terahertz light detection using fast terahertz quantum well photodetectors," Sci. Rep. **7**, 3452 (2017).
17. P. D. Grant, S. R. Laframboise, R. Dudek, M. Graf, A. Bezinger, and H. C. Liu, "Terahertz free space communications demonstration with quantum cascade laser and quantum well photodetector," Electron. Lett. **45**, 952–954 (2009).
18. C. Wang, J. Cao, L. Gu, Q. Wu, and Z. Tan, "20 Mbps wireless communication demonstration using terahertz quantum devices," Chinese Opt. Lett. Vol. 13, Issue 8, pp. 081402- **13**, 81402 (2015).
19. A. Vardi, N. Kheirodin, L. Nevou, H. Machhadani, L. Vivien, P. Crozat, M. Tchernycheva, R. Colombelli, F. H. Julien, F. Guillot, C. Bougerol, E. Monroy, S. Schacham, and G. Bahir, "High-speed operation of GaN/AlGaN quantum cascade detectors at λ≈1.55 μm," Appl. Phys. Lett. **93**, 193509 (2008).
20. R. Zhang, X. G. Guo, C. Y. Song, M. Buchanan, Z. R. Wasilewski, J. C. Cao, and H. C. Liu, "Metal-grating-coupled terahertz quantum-well photodetectors," IEEE Electron Device Lett. **32**, 659–661 (2011).
21. D. Palaferri, Y. Todorov, Y. N. Chen, J. Madeo, a. Vasanelli, L. H. Li, a. G. Davies, E. H. Linfield, and C. Sirtori, "Patch antenna terahertz photodetectors," Appl. Phys. Lett. **106**, 161102 (2015).
22. A. Benz, M. Krall, S. Schwarz, D. Dietze, H. Detz, A. M. Andrews, W. Schrenk, G. Strasser, and K. Unterrainer, "Resonant metamaterial detectors based on THz quantum-cascade structures," Sci. Rep. **4**, 4269 (2014).
23. K. Fan, A. C. Strikwerda, X. Zhang, and R. D. Averitt, "Three-dimensional broadband tunable terahertz metamaterials," Phys. Rev. B **87**, 161104 (2013).
24. B. Paulillo, J. M. Manceau, a. Degiron, N. Zerounian, G. Beaudoin, I. Sagnes, and R. Colombelli, "Circuit-tunable sub-wavelength THz resonators: hybridizing optical cavities and loop antennas," Opt. Express **22**, 21302 (2014).
25. B. Paulillo, J.-M. Manceau, L. H. Li, A. G. Davies, E. H. Linfield, and R. Colombelli, "Room temperature strong light-matter coupling in three dimensional terahertz meta-atoms," Appl. Phys. Lett. **108**, 101101 (2016).
26. C. Genet and T. W. Ebbesen, "Light in tiny holes," Nature **445**, 39–46 (2007).
27. A. E. Miroshnichenko, S. Flach, and Y. S. Kivshar, "Fano resonances in nanoscale structures," Rev. Mod. Phys. **82**, 2257–2298 (2010).
28. G. Xu, L. Li, N. Isac, Y. Halioua, a. Giles Davies, E. H. Linfield, and R. Colombelli, "Surface-emitting terahertz quantum cascade lasers with continuous-wave power in the tens of milliwatt range," Appl. Phys. Lett. **104**, 1–5 (2014).
29. H. C. Liu, C. Y. Song, A. J. SpringThorpe, and J. C. Cao, "Terahertz quantum-well photodetector," Appl. Phys. Lett. **84**, 4068–4070 (2004).
30. H. C. Liu, G. E. Jenkins, E. R. Brown, K. A. McIntosh, K. B. Nichols, and M. J. Manfra, "Optical heterodyne detection and microwave rectification up to 26 GHz using quantum well infrared photodetectors," IEEE Electron Device Lett. **16**, 253–255 (1995).
31. D. Bachmann, M. Rösch, M. J. Süess, M. Beck, K. Unterrainer, J. Darmo, J. Faist, and G. Scalari, "Short pulse generation and mode control of broadband terahertz quantum cascade lasers," Optica **3**, 1087 (2016).
32. D. M. Pozar, *Microwave Engineering, 4th Edition* (Wiley, 2011).

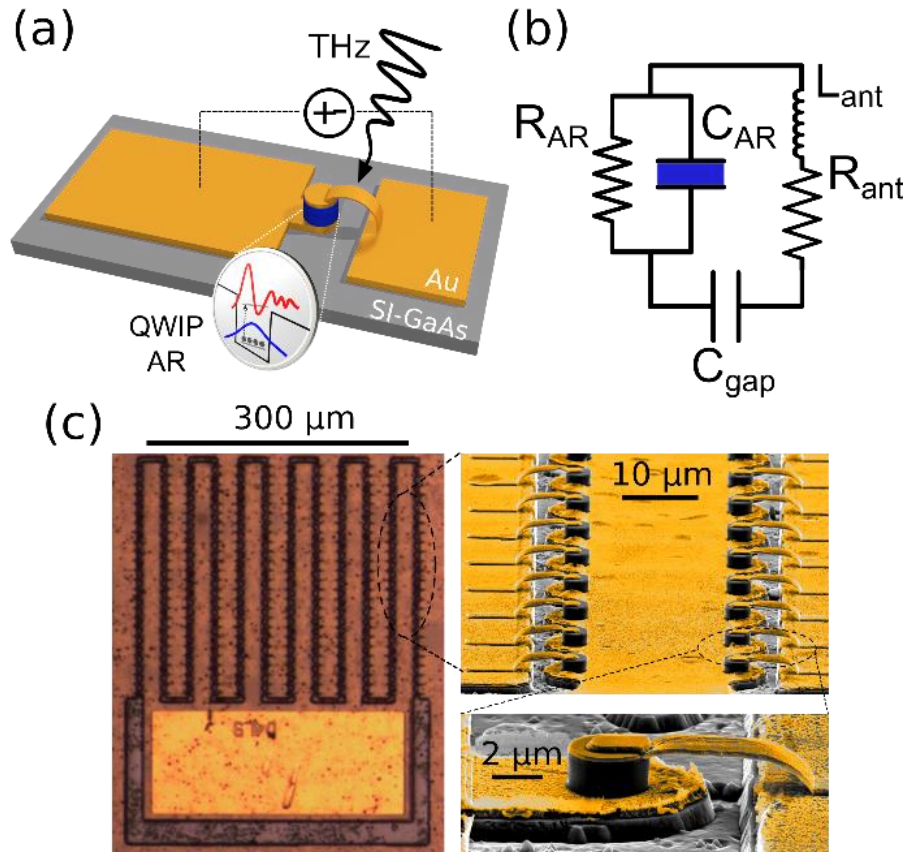

Fig. 1(a) Schematic diagram of the sub-wavelength meta-atom detector (single pixel). (b) RLC equivalent circuit: the meta-atom LC resonance is set by the MSM capacitance ($C_{AR}$) and the antenna inductance ($L_{ant}$). $R_{AR}$ and $R_{ant}$ are the active region and antenna resistances, respectively. $C_{gap}$ is the capacitance arising from the opening in the gold ground plane. (c) Pictures of the fabricated sample, at increasing magnification: optical microscopy image of the meta-atoms arranged in array geometry (left panel); SEM picture of the same array (right top panel); and, close view of a single meta-atom (right bottom panel). The SEM images have been colored to highlight the metallic regions.

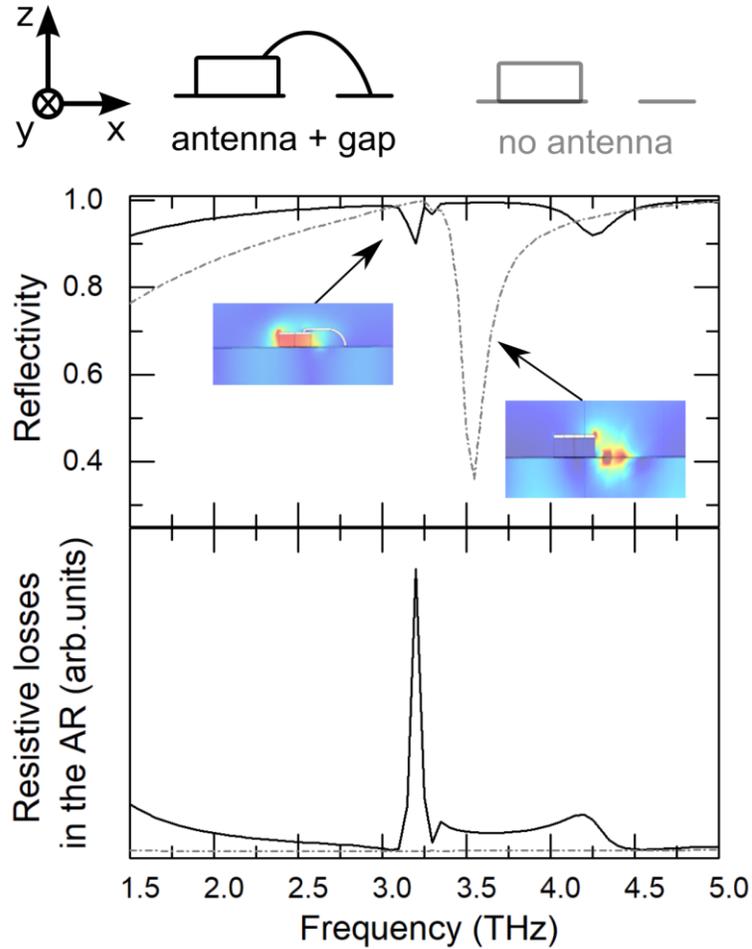

Fig. 2 Simulated optical response of an array of meta-atoms illuminated by normally incident THz radiation. (Top panel) Reflectivity spectra: when the antenna is not present and the MSM capacitor alone is considered (grey dot-dashed line), the spectrum shows a grating mode due to the sub-wavelength openings in the ground plane; when the antenna is added (black full line), the meta-atom LC resonance appears. (Bottom panel) Ohmic dissipation computed in the MSM cavity as a function of the excitation frequency revealing that energy is resonantly dissipated in the semiconductor only if the inductive antenna completes the RLC circuit. [see Supplementary Material for further details].

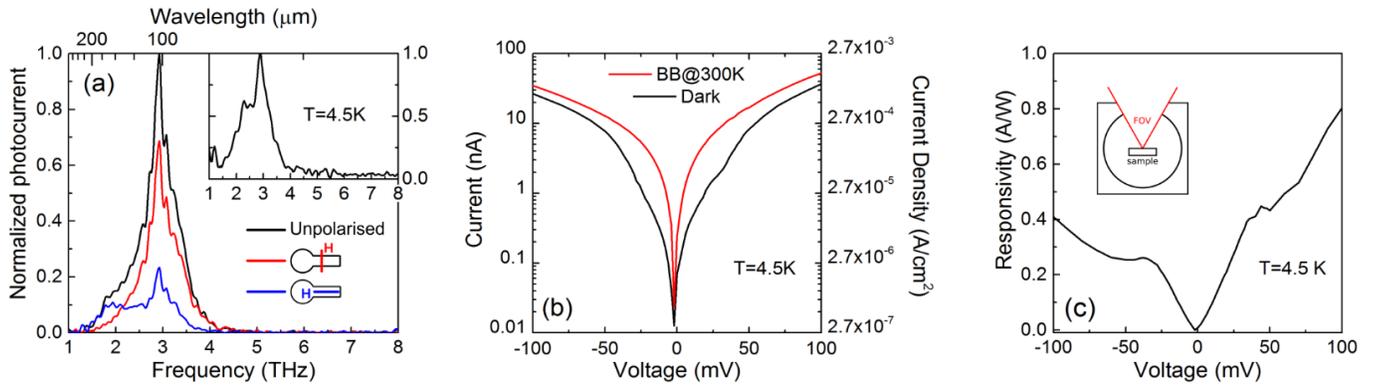

Fig. 3(a) Photo-current spectra from an array of 300 meta-atom detectors for unpolarized (black curve) and polarized (red and blue curves) radiation from a Hg lamp: THz radiation is fed into the device when the magnetic field is orthogonal to the antennas.
(b) Typical current-voltage curves under dark conditions (black solid line) and under illumination from a 300 K black body (red solid line) for an array of meta-atom detectors. (c) Estimated array responsivity. The inset clarifies the definition of the field of view (FOV) of the detector inside the cryostat. All the measurements are performed at 4.5 K and using a TPX cryostat window (no other filters have been added).

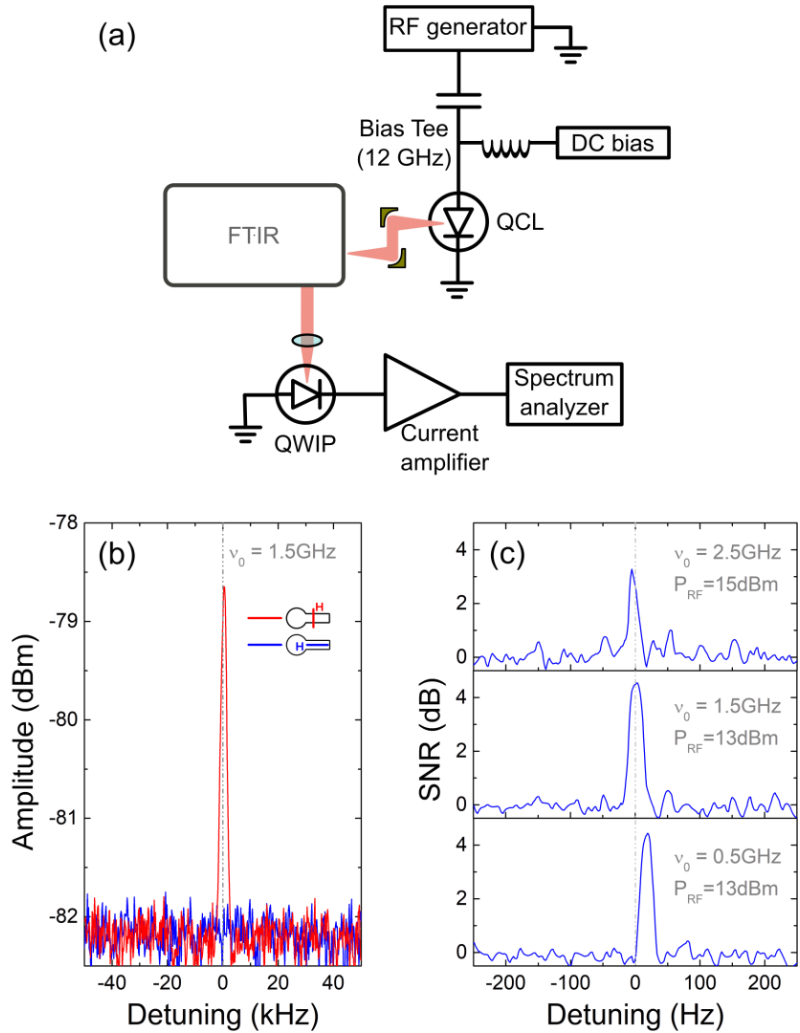

Fig. 4 Optical detection of ultrafast THz signals with a meta-atom array. (a) Sketch of the experimental arrangement: the emitter is an RF-modulated THz QCL operating in continuous wave (f=3 THz, I=220 mA, T=10 K). The receiver is an unbiased meta-atom array (T=5 K), and the RF signal is recovered on a spectrum analyzer after an amplification stage. (b) RF spectrum measured when the QCL is modulated at 1.5 GHz ($P_{RF}$=13 dBm, 30 dB amplification). Two polarizations are shown: magnetic field orthogonal (red curve) and parallel (blue curve) to the antennas. The latter is taken as a noise-reference. (c) Normalized RF spectra acquired when the QCL is driven at 0.5, 1.5 and 2.5 GHz (20 dB amplification).

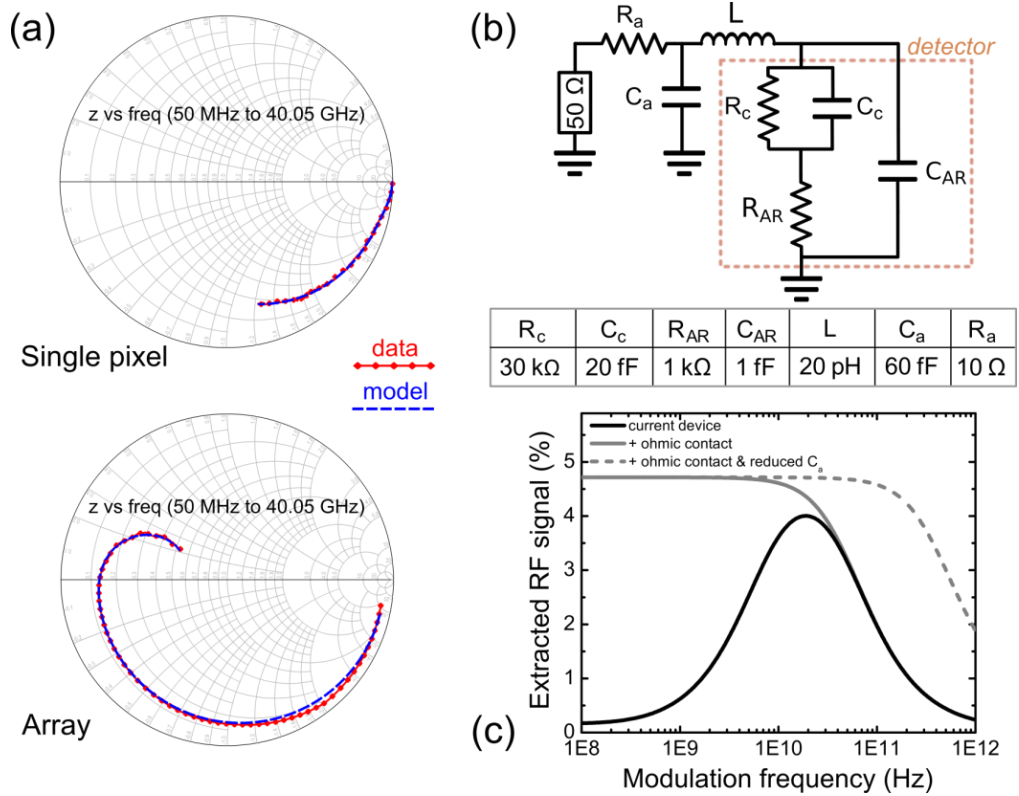

Fig. 5: (a) Normalized detector impedance $z=Z/Z_0$ as a function of frequency for a single meta-atom and an array (T=300K): experimental (red symbols) vs equivalent circuit model (blue dashed line). (b) Equivalent circuit of a single meta-atom detector, including the access contact lines. (c) Extracted RF signal $I_{out}/I_{in}$ (simulation) as a function of frequency for: (i) a single detector in the actual geometry (full black line), (ii) a single detector with top ohmic contact (full grey line), (iii) a single detector with top ohmic contact and reduced contact pad capacitance (dashed grey line).